\documentclass[a4paper]{jpconf}
\usepackage{cite}
\usepackage{enumerate}   
\usepackage{amsfonts,amsmath,amssymb,latexsym,graphicx,xcolor,
cancel,textcomp,graphicx,
BOONDOX-calo}

\begin{document}

\title{Conjecture of perturbative QED breakdown \\ at $\boldsymbol{\alpha\chi^{2/3}\gtrsim 1}$}
\author{Alexander Fedotov}
\address{National Research Nuclear University ``MEPhI'' (Moscow Engineering Physics Institute),\\ 
Kashirskoe sh. 31, 115409 Moscow, Russian Federation}
\ead{am\_fedotov@mail.ru}
\begin{abstract}
Invited talk given at Symposium ``Extreme Light Technologies, Science, and Applications''  (LPHYS'16, Yerevan, 11-15 July 2016). The old Ritus-Narozhny conjecture of possible breakdown of Intense Field QED perturbation theory for ultrarelativistic particles passing transversely through a strong electromagnetic field is reviewed with a special emphasis on its possible significance for near-future experiments.
\end{abstract}

\begin{flushright}
\parbox{0.7\textwidth}{\it Dedicated to memory of Nikolay Borisovich  Narozhny (1940-2016)}
\end{flushright}

\section{Introduction}
Professor Nikolay Borisovich Narozhny, the eminent Russian theoretician and the organizer of the Symposium ``Extreme Light Technologies, Science, and Applications'', passed away on 15 February 2016 \cite{ICUIL, UFNsite}. I was lucky to work under his supervision on PhD thesises in 1997--2000 and to remain one of his main coauthors since then. Most of his pioneer works are well recognized by the community, e.g. calculation of  probabilities for photon emission and pair photoproduction in circularly polarized electromagnetic wave \cite{narozhnyi1964quantum},  first calculation of  polarization operator in a constant crossed field \cite{narozhny1969propagation}, first direct calculation of  spontaneous pair production in electric field \cite{narozhny1970simplest}, or the effect of collapses and revivals in cavity QED \cite{eberly1980periodic,narozhny1981coherence}.

However, his probably the most deep and significant contribution (or at least claimed as such in his Dr.Sc. dissertation back in 1982), the $\alpha^3$-order IFQED calculations \cite{narozhny1979radiation,narozhny1980expansion,morozov1981} proving the original Ritus conjecture \cite{ritus1970radiative} of possible  breakdown of perturbative QED at $\alpha\chi^{2/3}\gtrsim 1$, still remains rather unknown. In this talk I am going to give the review of that old idea. My particular  aims are: to explain some known arguments in favor of the conjecture, to give some insights into its meaning, and to stress its significance for the near future progress of laser-matter interaction studies at extreme intensities.

\section{Radiation corrections in Classical Electrodynamics and ordinary QED}

Let me start with brief reminder of the basics of radiation corrections issues in Classical Electrodynamics and in ordinary QED. In Classical Electrodynamics the self-energy of a particle at rest reads
\begin{equation*}
\mathcal{E}_\text{em}=\frac12\int\limits d^3r\int\limits d^3r'\,\frac{\rho(\vec{r})\rho(\vec{r}')}{|\vec{r}-\vec{r}'|}
\simeq \frac{e^2}{r_0},
\end{equation*}
and becomes $\gtrsim mc^2$ at $r_0\lesssim r_e\equiv\frac{e^2}{mc^2}$. The length $r_e$ is traditionally called the `classical electron radius'. The radiation reaction force (in proper reference frame `p')
\begin{equation*}
F_\text{rad}=\frac23\frac{e^4}{m^2c^4}E_\text{p}^2,
\end{equation*}
produces over the distance $r_e$ the work
\begin{equation*}
A=F_\text{rad}r_e\simeq \frac{e^6E_\text{p}^2}{m^3c^6},
\end{equation*}
which also $\gtrsim mc^2$ at $E_\text{p}\gtrsim E_\text{cr}\equiv \frac{m^2c^4}{e^3}$. The distance $r_e$ and the field strength $E_\text{cr}$ are considered as limits of applicability of Classical Electrodynamics. 

Things are changed drastically in QED, though. To simplify formulas, let us consider scalar QED (sQED) as an example. The free quantized scalar charged field reads
\begin{equation*}
\hat{\Psi}(x)=\sum\limits_{\vec{p}}\frac1{\sqrt{2V\varepsilon_{\vec{p}}}}\left(e^{-ipx}\hat{a}_{\vec{p}}+e^{ipx}\hat{b}_{\vec{p}}^\dagger\right),
\end{equation*}
where $\hat{a}_{\vec{p}}$ and $\hat{b}_{\vec{p}}$ are charged particle and antiparticle annihilation operators, respectively. The sQED $4$-current operator
\begin{multline*}
\hat{j}_\mu(x)=ie\,:\hat{\Psi}^\dagger(x)\stackrel{\leftrightarrow}{\partial_\mu}\hat{\Psi}(x):\,=\sum\limits_{\vec{p},\vec{p'}}\frac{e}{2V\sqrt{\varepsilon_{\vec{p}}\varepsilon_{\vec{p}'}}}\left\{\left(p_\mu+p'_\mu\right)\left[\underbrace{e^{i(p'-p)x}\hat{a}_{\vec{p}'}^\dagger\hat{a}_{\vec{p}}}_{\text{particle current}}-\underbrace{e^{-i(p'-p)x}\hat{b}_{\vec{p}}^\dagger\hat{b}_{\vec{p}'}}_{\text{antiparticle current}}\right]+\right.
\\\left.+
\left(p_\mu-p'_\mu\right)\left[\underbrace{e^{-i(p'+p)x}{\color{black}\hat{b}_{\vec{p}'}\hat{a}_{\vec{p}}}-e^{i(p'+p)x}{\color{black}\hat{a}_{\vec{p}'}^\dagger\hat{b}_{\vec{p}}^\dagger}}_{\text{non-diagonal terms}}\right],
\right\}
\end{multline*}
contains a non-diagonal part, which corresponds to annihilation and creation of a virtual pair.

By passing to collective coordinate $\vec{R}=\frac{\vec{r}+\vec{r}'}{2}$ and separation $\vec{\xi}=\vec{r}-\vec{r}'$, the self-energy can be cast into form
\begin{equation*}
\mathcal{E}_\text{em}=\frac12\int\limits d^3\xi\,\frac{C(\vec{\xi})}{|\vec{\xi}|},
\end{equation*}
where $C(\vec{\xi})$ is difference between the expectation values of the charge density correlator in 1-particle state and in vacuum:
\begin{multline*}
C(\vec{\xi})=\int\limits d^3R\,\langle 1_\text{rest}|\hat{j}^0\left(\vec{R}+\frac{\vec{\xi}}2\right)\hat{j}^0\left(\vec{R}-\frac{\vec{\xi}}2\right)-:\quad:|1_\text{rest}\rangle =\\
=\frac{e^2}2\int\limits \frac{d^3p}{(2\pi)^3} \underbrace{\left(1+\frac{m}{{\varepsilon_{\vec{p}}}}\right)\,e^{i\vec{p}\vec{\xi}}}_\text{from $\hat{a}^\dagger\hat{a}\hat{a}^\dagger\hat{a}$}-\frac{e^2}2\int\limits \frac{d^3p}{(2\pi)^3}\underbrace{\left(1-\frac{m}{{\varepsilon_{\vec{p}}}}\right)\,e^{i\vec{p}\vec{\xi}}}_\text{from $\hat{b}\hat{a}^\dagger\hat{a}\hat{b}^\dagger$}.
\end{multline*}
In classical limit ($m\to\infty$) the first (particle) contribution in braces reduces (as expected) to $e^2\delta^{(3)}(\vec{\xi})$, while the second term vanishes. This results in linear divergency as above: $\mathcal{E}_\text{em}\propto \frac{e^2}2\int\limits d^3\xi\frac{\delta^{(3)}(\vec{\xi})}{|\vec{\xi}|}\simeq\frac{e^2}{r_0}$, $r_0\to 0$. 

However, in general setting the leading divergency is canceled by the virtual pairs contribution \cite{weisskopf1939self}:
\begin{multline*}
C(\vec{\xi})=\frac{e^2}2\left(\cancel{\delta^{(3)}(\vec{\xi})}-\frac{m^2}{2\pi^2|\vec{\xi}|}K_1(m |\vec{\xi}|)\right)-\frac{e^2}2\left(\cancel{\delta^{(3)}(\vec{\xi})}+\frac{m^2}{2\pi^2|\vec{\xi}|}K_1(m |\vec{\xi}|)\right)=\\=-\frac{e^2 m^2}{2\pi^2|\vec{\xi}|}K_1(m |\vec{\xi}|)\simeq \frac{e^2 m}{\pi^2|\vec{\xi}|^2},\quad \vec{\xi}\to 0,
\end{multline*}
so that
\begin{equation*}
\mathcal{E}_\text{em}\simeq \frac{e^2m}{\pi^2}\int\limits \frac{d^3\xi}{|\vec{\xi}|^3}\propto e^2m\log\left(\frac1{m r_0}\right),\quad r_0\to 0.
\end{equation*}
Thus, a pointlike charge is replaced by a cloud of virtual pairs of Compton size
$\simeq l_C=\frac1{m}\simeq 137 r_e$. Divergency is still present but now gets much weaker (logarithmic vs linear) than in Classical Electrodynamics. After renormalization (which is all the same required for physical reasons, albeit $\mathcal{E}_\text{em}\simeq \alpha m\log\left(\frac1{m r_0}\right)\ll m$ for any reasonable value of $r_0$), the coupling constant becomes effectively `running', and its energy dependence essentially mimics the character of divergency: $\alpha(\varepsilon)\simeq \alpha\log\left(\frac{\varepsilon}{m}\right)$, $\varepsilon\gg m$ (high energy `stripping'). Note that $\alpha(\varepsilon)$ nevertheless remains small for all reasonable values of energy. Review and classification of the variety of high-energy QED processes\cite{Gorshkov1973,baier1981inelastic} demonstrates that all the cross sections remain also small $\sigma(\varepsilon)\lesssim \alpha^n r_e^2 \log^k\left(\frac{\varepsilon}{m}\right)$ within all the reasonable energy range. Thus perturbation theory in ordinary QED works extremely well within all the reasonable range of parameters.

\section{Radiation corrections in Intense Field QED}

However, in external field with $a_0\equiv \frac{e\sqrt{-A_\mu^2}}{m}\gg 1$ (for optical lasers at intensity $I_L\gtrsim 10^{18}$W/cm$^2$)perturbation theory \emph{with respect to interaction with that field} breaks down and all-order summation is needed, which reduces only to replacement of free external lines and propagators by the exact (`dressed') ones in external field:
\begin{center}
\includegraphics[width=0.9\textwidth]{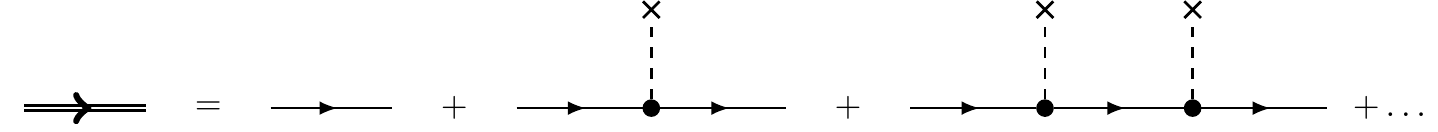}
\end{center}
For several cases (including the most important paradigmatic case of  constant crossed field, which corresponds to $a_0\gg1$ and relativistic motion across the field), the equation 
\begin{center}
\includegraphics[width=0.6\textwidth]{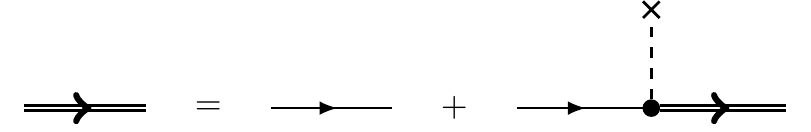}
\end{center}
can be solved in closed form.
Note that in CCF electrons / photons are characterized by a single Lorentz- and gauge-invariant parameter $\chi=\frac{e}{m^3}\sqrt{-(F_{\mu\nu}p^\nu)^2}$ / $\varkappa=\frac{e}{m^3}\sqrt{-(F_{\mu\nu}k^\nu)^2}$ - for electron this is just its proper acceleration in Compton units.

It was noticed already at its birth \cite{nikishov1964quantum,narozhny1969propagation,ritus1970radiative} that  in IFQED radiation corrections are growing unusually fast with $\chi$ or $\varkappa$ (i.e. with both energy and field strength):
\begin{multline*}
M^{(2)}(\chi)=\raisebox{-0.25\height}{\includegraphics[width=2cm]{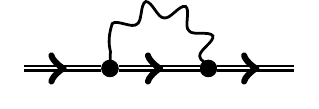}}\simeq \alpha m\chi^{2/3},\quad 
W_{e^{\pm}\to e^{\pm}\gamma}(\chi)=\frac{2m}{p_0}\,{\rm Im}\,M^{(2)}\simeq \frac{\alpha m^2}{p_0}\chi^{2/3},\quad \chi\gg 1;
\end{multline*}
\begin{multline*}
\mathcal{P}^{(2)}(\varkappa)=\raisebox{-0.45\height}{\includegraphics[width=2cm]{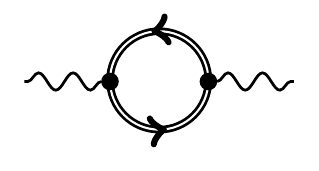}}\simeq \alpha m^2\varkappa^{2/3},\quad 
W_{\gamma\to e^+e^-}(\varkappa)=\frac2{k_0}\,{\rm Im}\,\mathcal{P}^{(2)}\simeq \frac{\alpha m^2}{k_0}\varkappa^{2/3},\quad \varkappa\gg 1;
\end{multline*}

This implies that for $\chi,\,\varkappa\gtrsim \alpha^{-3/2}\simeq 1.6\times 10^3$ (or, equivalently, $E_\text{p}\gtrsim 12 E_\text{cr}\simeq 1.6\times 10^3 E_S$)
\begin{equation*}
M^{(2)}\simeq m,\quad \mathcal{P}^{(2)}\simeq m^2
\end{equation*}
and that in proper reference frame
\begin{equation*}
t_e\sim W_{e^{\pm}\to e^{\pm}\gamma}^{-1}\simeq t_C,\quad 
t_\gamma\sim W_{\gamma\to e^+e^-}^{-1}\simeq t_C
\end{equation*}
These means that radiation corrections become not small and radiation-free motion could show up only at Compton scale (where localization is all the same impossible).
 
\begin{center}
\begin{table}[t]
\centering
\caption{\label{tab1}Typical values of electron energy ($\varepsilon_\text{in}$) and laser intensity ($I_L$) required for attaining the regime $\alpha\chi^{2/3}\simeq 1$ in counterpropagating setup.} 
\setlength{\tabcolsep}{7pt}\renewcommand{\arraystretch}{1.5}
\begin{tabular}{|l|c|c|c|c|}
\hline
$\varepsilon_\text{in}=m\gamma_\text{in}$, GeV &  $800$&  $80$&  $8$&  $0.8$\\ \hline
$E/E_S$ & $10^{-3}$ & $10^{-2}$ & $0.1$  & $1$ \\ \hline
$I_L$,W/cm$^2$ & $5\times10^{23}$ & $5\times10^{25}$ & 
$5\times10^{27}$ & $5\times10^{29}$ \\ \hline
\end{tabular}
\end{table}
\end{center}

For high-energy electrons counterpropagating a laser pulse we have $\chi\sim \frac{E\gamma_\text{in}}{E_S}$. Some typical values of parameters required to fulfill the condition $\alpha\chi^{2/3}\gtrsim 1$ are listed in Table~\ref{tab1}.
Observe that this threshold could be almost overcome experimentally by combining state-of-the-art laser systems with the future ILC--class TeV lepton colliders.
It is also worth noting that the table assumes transverse propagation across the field. For self-sustained (A-type) cascades \cite{fedotov2010limitations,elkina2011qed} $E\gtrsim \alpha E_S$ and
\begin{multline*}
\measuredangle (\vec{p},\vec{E})\sim \left(\frac{\alpha E_S}{E}\right)^{1/4}\lesssim 1,\quad\chi\sim \left(\frac{E}{\alpha E_S}\right)^{3/2}\gtrsim 1,\quad 
\text{but}\quad \alpha\chi^{2/3}\sim \frac{E}{E_S}\ll 1
\end{multline*}

During 1972--1981 some higher-order radiation corrections in IFQED were either obtained or estimated by Ritus, Narozhny and Morozov \cite{ritus1972radiative,morozov1975elastic,morozov1977elastic,narozhny1979radiation,narozhny1980expansion}. 
Namely, for mass corrections it was obtained:
\begin{multline*}
\mbox{\LARGE $\frac{M}{m} =$}\underbrace{\raisebox{-0.25\height}{\includegraphics[width=2cm]{mass2.pdf}}}_{{\color{black}\simeq\alpha\chi^{2/3}\;(\text{Ritus, 1970 \cite{ritus1970radiative}})}}+\underbrace{\raisebox{-0.1\height}{\includegraphics[width=3cm]{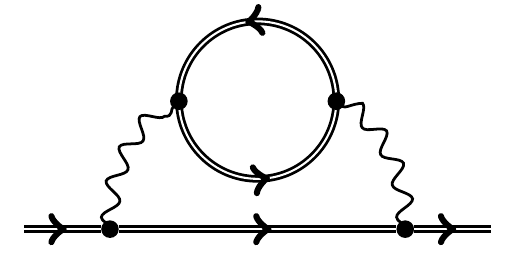}}}_{{\color{black}\simeq\alpha^2\chi\log{\chi}\;(\text{Ritus, 1972 \cite{ritus1972radiative}})}}+\underbrace{\includegraphics[width=3cm]{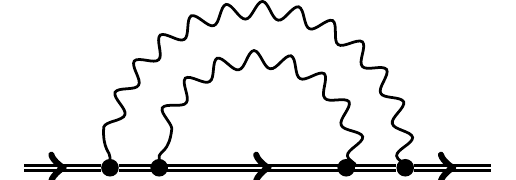}}_{\simeq\alpha^2\chi^{2/3}\log{\chi}\;(\text{Morozov\&Ritus, 1975 \cite{morozov1975elastic}})}\\
+\quad\underbrace{\raisebox{-0.45\height}{\includegraphics[width=3cm]{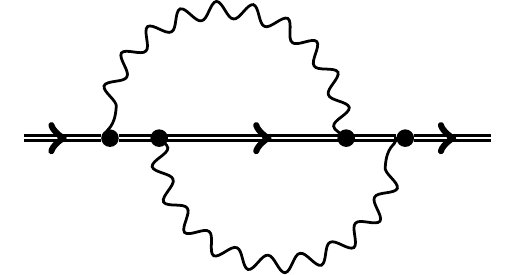}}}_{{\color{black}\simeq\alpha^2\chi^{2/3}\log{\chi}\;(\text{?})}}+\underbrace{\includegraphics[width=3cm]{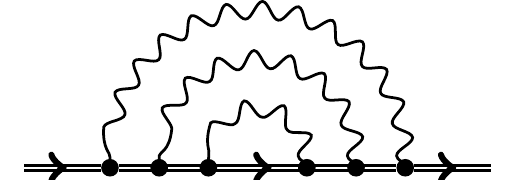}}_{\simeq\alpha^3\chi^{2/3}\log^2{\chi}\;(\text{Narozhny, 1979 \cite{narozhny1979radiation}})}+\underbrace{\raisebox{-0.12\height}{\includegraphics[width=3cm]{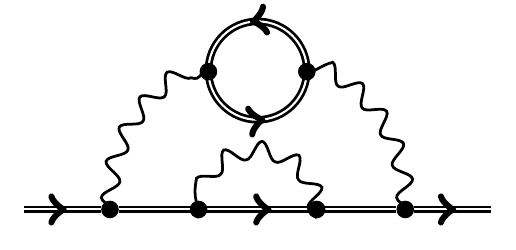}}}_{\simeq\alpha^3\chi^{4/3}\;(\text{Narozhny, 1979 \cite{narozhny1979radiation}})}\\ 
+\underbrace{\includegraphics[width=3cm]{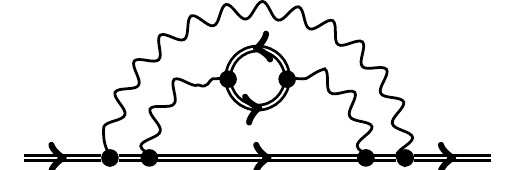}}_{\simeq\alpha^3 \chi\log^2{\chi}\;(\text{Narozhny, 1980 \cite{narozhny1980expansion}})}+\underbrace{\raisebox{-0.15\height}{\includegraphics[width=3cm]{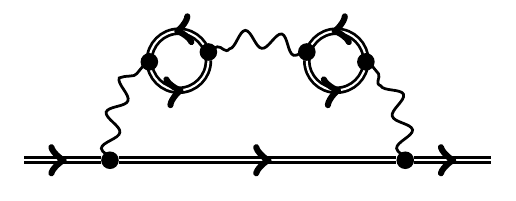}}}_{\boxed{\scriptstyle{\color{black}\simeq\alpha^3\chi^{5/3}\;(\text{Narozhny, 1980 \cite{narozhny1980expansion}})}}}+\underbrace{\raisebox{-0.1\height}{\includegraphics[width=3cm]{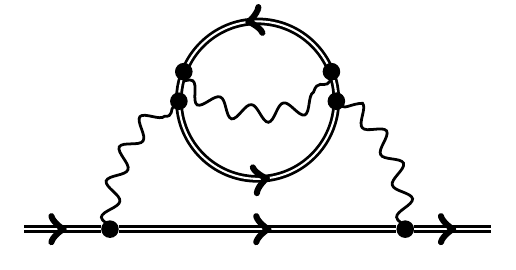}}}_{{\color{black}\simeq\alpha^3 \chi^{2/3}\log^2{\chi}\;(\text{?})}}\\+\quad\underbrace{\includegraphics[width=3cm]{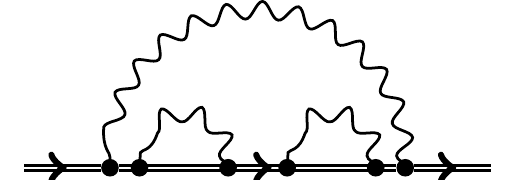}}_{{\color{black}\simeq\alpha^3\chi^{2/3}\log^2{\chi}\;(\text{?})}}\quad+\ldots
\end{multline*}
(here and below the contributions of the diagrams shown with question mark were either calculated in draft but unpublished, or even guessed basing on their infrared behavior). For polarization correction it was obtained:
\begin{multline*}
\mbox{\LARGE $\frac{\mathcal{P}}{m^2} =$}
\underbrace{\raisebox{-0.45\height}{\includegraphics[width=2cm]{polarization2.pdf}}}_{\stackrel{{\color{black}\scriptstyle\simeq\alpha\varkappa^{2/3}}}{{\color{black}(\text{Narozhny, 1968 \cite{narozhny1969propagation}})}}}
+\underbrace{\raisebox{-0.5\height}{\includegraphics[width=2.5cm]{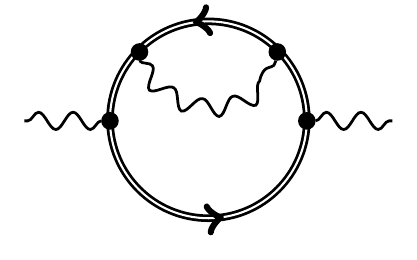}}}_{\stackrel{{\color{black}\scriptstyle\simeq\alpha^2\varkappa^{2/3}\log{\varkappa}}}{{\color{black}(\text{Morozov\&Narozhny, 1977 \cite{morozov1977elastic}})}}}
+\quad\underbrace{\raisebox{-0.5\height}{\includegraphics[width=2.5cm]{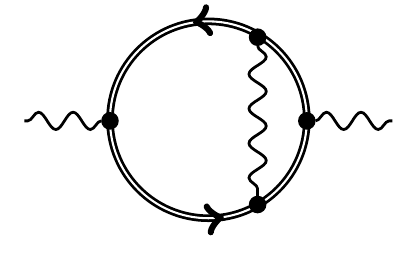}}}_{{\color{black}\simeq\alpha^2\varkappa^{2/3}\log{\varkappa}\;\text{(?)}}}
+\underbrace{\raisebox{-0.5\height}{\includegraphics[width=2.5cm]{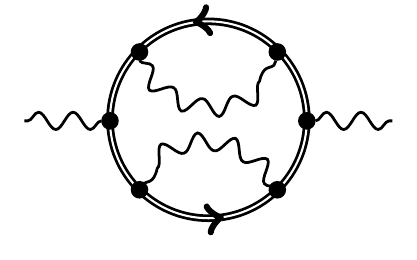}}}_{\stackrel{\scriptstyle\simeq\alpha^3\varkappa^{2/3}\log{\varkappa}}{\text{(Narozhny, 1979 \cite{narozhny1979radiation})}}}+\\
+\underbrace{\raisebox{-0.5\height}{\includegraphics[width=2.5cm]{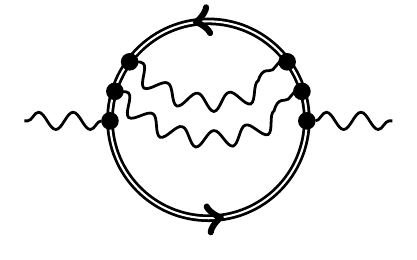}}}_{\stackrel{\scriptstyle\simeq\alpha^3\varkappa^{2/3}\log{\varkappa}}{\text{(Narozhny, 1979 \cite{narozhny1979radiation})}}}+\underbrace{\raisebox{-0.5\height}{\includegraphics[width=2.5cm]{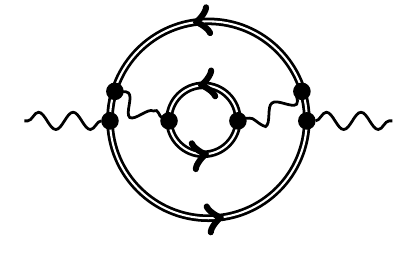}}}_{\boxed{\stackrel{{\color{black}\scriptstyle\simeq\alpha^3\varkappa\log^2{\varkappa}}}{\scriptstyle\text{\color{black}(Narozhny, 1980 \cite{narozhny1980expansion})}}}}
+\underbrace{\raisebox{-0.5\height}{\includegraphics[width=2.5cm]{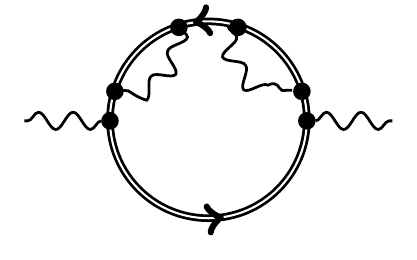}}}_{{\color{black}\simeq\alpha^3\varkappa^{2/3}\log^2{\varkappa}\;\text{(?)}}}
+\underbrace{\raisebox{-0.5\height}{\includegraphics[width=2.5cm]{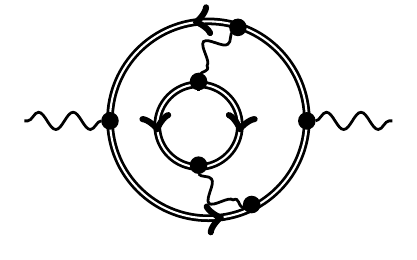}}}_{{\color{black}\simeq\alpha^3\varkappa^{2/3}\log^2{\varkappa}\;\text{(?)}}}+\ldots
\end{multline*}
One can observe that within each order different types of diagrams depend on $\chi$ or $\varkappa$ rather differently: some higher-order diagrams acquire just additional logarithmic factors as in ordinary QED, but some others reveal power dependence, with power growing in higher orders. In fact, when Ritus first obtained his $e^4$-order polarization correction to mass operator \cite{ritus1972radiative}, he initially suggested $\alpha\chi^{1/3}$ as roughly an expansion parameter in IFQED perturbation theory (which should be also puzzling, in fact). The main goal of Narozhny \cite{narozhny1980expansion} was demonstrating that the $e^6$-order two-loop polarization diagram grows even faster and that the true expansion parameter should be $\alpha\chi^{2/3}$ instead. Moreover, he argued that the $4$-th order contribution to the mass operator and $2$-nd and $4$-th order contributions to polarization operator are just occasionally suppressed and that corrections to mass operator outstrip those to polarization operator by one order, so that the same parameter $\alpha\chi^{2/3}$ should show up in polarization operator in $8$-th order as well.

Basing on these results, one could get an impression that the leading contributions probably originate in diagrams containing maximal number of successive uncorrected polarization loops (if true, it could drastically simplify further considerations). However, the only analyzed vertex correction \cite{morozov1981}
\begin{multline*}
\mbox{\LARGE $\frac{\Gamma}{e}=$}\underbrace{\raisebox{-0.8\height}{\includegraphics[height=1cm]{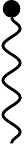}}}_{=i\gamma^\mu}+\underbrace{\raisebox{-0.42\height}{\includegraphics[width=2cm]{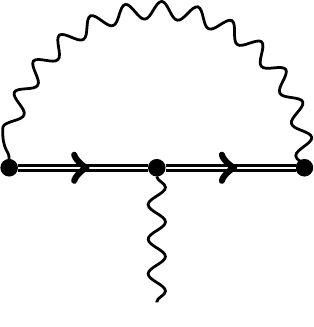}}}_{\boxed{\scriptstyle{\color{black}\simeq\alpha \chi^{2/3}\;(\text{Morozov,Narozhny\&Ritus,1981 \cite{morozov1981}})}}}+\ldots\hfil\hfil\hfil
\end{multline*}
also grows as $\alpha\chi^{2/3}$, so that some further calculations are definitely needed to support this idea. According to my knowledge, no further progress was made on these calculations since 1981.

\section{Qualitative estimates and physical meaning}

On defense, Nikolay Borisovich was asked for a `simple words' physical reasoning (interpretation) for appearance of the parameter $\alpha\varkappa^{2/3}$. And his answer was about that in ultrarelativistic case $\mathcal{P}=\alpha m^2 F(\varkappa)$ should not depend on $m$. Then, since $\varkappa\propto m^{-3}$, it should be $F(\varkappa)\propto \varkappa^{2/3}$ unambiguously for $\varkappa\gg 1$. Unfortunately, this argument doesn't work for $M$, $\Gamma$ and higher orders.

However, recently a more visual and direct explanation was seemingly found\cite{fedotov2015qualitative}. Consider for definiteness\footnote{Ultra\-relativistic kinematics is in fact similar for all the processes.} formation time $t$ and length $l$ for the polarization operator $\mathcal{P}^{(2)}(\varkappa\gg 1)$. Assuming $k$ and $p$ large and also that initially $\vec{k}\perp\vec{E}$, the energy uncertainty of the virtual process $\gamma\to e^-e^+\to\gamma$
\begin{multline*}
\Delta\varepsilon(t)=\sqrt{p^2+e^2E^2t^2+m^2}+\sqrt{(k-p)^2+e^2E^2t^2+m^2}-k
\simeq \cancel{p}+\frac{e^2E^2t^2+m^2}{2p}+\\
+\cancel{k}-\cancel{p}+\frac{e^2E^2t^2+m^2}{2(k-p)}-\cancel{k}
=\frac{k\left(e^2E^2t^2+m^2\right)}{2p(k-p)}\ge \frac{2\left(e^2E^2t^2+m^2\right)}{k}
\end{multline*}
Assume additionally (to be confirmed by the result) that $eEt\gg m$. Then $\Delta\varepsilon(t)\simeq \frac{e^2E^2t^2}{k}$ and from the uncertainty principle $\Delta \varepsilon\cdot t\sim 1$ we obtain 
\begin{equation*}
t,l_\parallel\simeq \left(\frac{k}{e^2E^2}\right)^{1/3}\equiv \frac{k}{m^2\varkappa^{2/3}}\equiv \frac{m}{eE}\varkappa^{1/3},
\end{equation*}
where $\varkappa=eEk/m^3$. Note that $eEt\simeq m\varkappa^{1/3}\gg m$, exactly as assumed. It turns out that this key simple estimate is confirmed by direct derivation of effective formation region for the integrals that define the quantum amplitude\cite{morozov1975elastic}. 

Transverse separation $l_\perp\simeq \frac{eEt^2}{k}\simeq \frac1{m\varkappa^{1/3}}\equiv \frac1{(eEk)^{1/3}}$ (note it is $m$-independent). Maybe a bit counterintuitively, charge separation reduces (rather than increases) with the field -- this quantum effect arises because $t$ reduces too fast.
Moreover, strong ($\varkappa\gg 1$) field is capable for confining virtual pairs to distances smaller than $l_C=\frac1{m}$ (this is very reminiscent to the Ritus's observation \cite{ritus1975lagrangian} of  strong field--small distance correspondence)!
In `proper' reference frame\footnote{I.e. where the photon is `soft' ($k'\sim m$) and $E_P\sim\varkappa E_S$.} $l_\parallel'\sim \frac{m}{k}l_\parallel\sim \frac1{m\varkappa^{2/3}}\ll l_\perp$, thus $l_\parallel'$ is the smallest scale. Surprisingly, for $\alpha\chi^{2/3}\sim 1$ it coincides to the classical electron radius $r_e$!
Now polarization operator should be defined by these scales: $\mathcal{P}(\varkappa)\simeq e^2 /{l_\perp}^2(\varkappa)$. Similarly, $M\simeq e^2/ l_\parallel'(\varkappa)$ and the parameter $\alpha\chi^{2/3}\equiv \frac{e^2/l_\parallel'}{m}$ may be viewed as just Coulomb energy to rest energy ratio.

\section{Nonperturbative regime and cascades}

QED cascades may be in direct relation with the nonperturbative regime under discussion. Indeed, at $\chi,\varkappa\gg 1$ the work produced by the field during the formation time $t$, $A\simeq eE l_\perp\simeq \Delta\varepsilon$. This means that the virtual intermediate states in such a regime are in fact rather close to mass shell. This corresponds to the fact that at $\chi,\varkappa\gg 1$ the real and imaginary parts of the mass and polarization operators become of the same order. Moreover, according to the optical theorem, these imaginary parts define the probabilities of the processes described by diagrams that can be obtained by making cuts over intermediate states. By restricting to diagrams containing polarization corrections to propagators, i.e. those of type:
\begin{center}
\raisebox{-0.8\height}{\includegraphics[height=6cm]{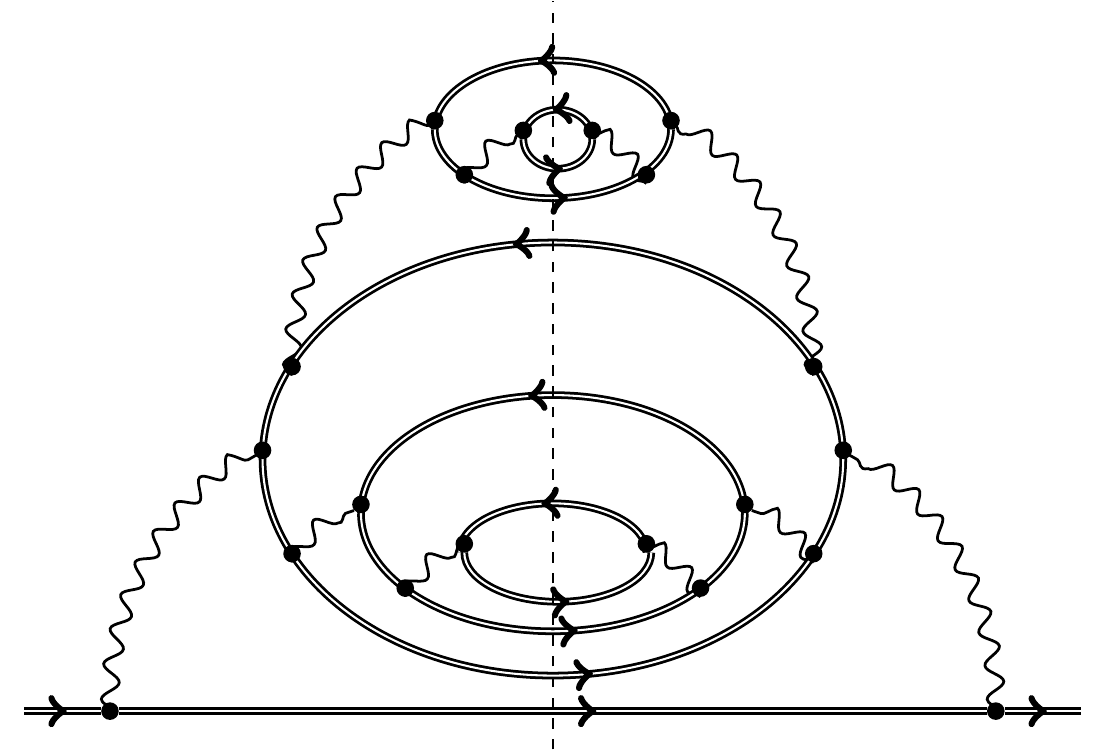}}
\end{center}
we obtain by such cuts exactly the diagrams for cascade processes (at that cutting corrections for mass operator corresponds to electron-seeded cascades, and for polarization operator to photon-seeded cascades). Intuitively, passing from perturbative to non-perturbative regime might correspond to transition from `ordinary' (S-type) to self-sustained (A-type) cascades \cite{mironov2014collapse}. As already mentioned, for self-sustained cascades $\alpha\chi^{2/3}\simeq 1$ corresponds to $E\simeq E_S$. 
According to our recent results \cite{fedotov2010limitations}, attaining of such field strengths can be questionable (notably due to cascade generation), provided the field is capable for spontaneous pair creation. This conclusion, however, is not related to fields that are incapable for that (e.g. plane wave field or CCF in which, on other hand, self-sustained cascades can not originate at all). Besides, our cascade model was developed by neglecting radiation corrections. Since development of self-sustained cascades necessarily requires field variation, possible significance of self-sustained cascades for non-perturbative dynamics yet remains unclear.

\section{Conclusion}
The conjecture that radiation corrections in IFQED are growing as a power of energy and field strength is really puzzling and challenging for theoreticians:
\begin{itemize}
\item In such a regime QED may become a truly non-perturbative theory: all the numerous results published by now may become invalid!
\item In particular, the whole IFQED approach we got used to, should also break down, as the external field lines used from the very beginning in `exact' propagators 
\begin{center}
\includegraphics[width=0.9\textwidth]{exact_prop.pdf}
\end{center}
should be radiatively corrected as well! 
\item Possible hints: for $\alpha\chi^{2/3}\sim 1$:
\begin{enumerate}[(i)]
\item At least for mass operator, the known leading radiation corrections arise from diagrams containing maximal number of successive uncorrected polarization loops;
\item $l_\parallel'\simeq r_e$;
\item Possible relation with self-sustained QED cascades at $E\simeq E_S$.
\end{enumerate}
\end{itemize}
Importantly, the regime $\alpha\chi^{2/3}\gtrsim 1$ may `soon' appear observable for experimentalists. But unfortunately, potential significance of the conjecture has still been underestimated by the community.

\section*{Acknowledgements}
The work was supported by RFBR grant 16-02-00963a and the Competitiveness Enhancement Program of NRNU MEPhI. I am grateful to A. Ilderton, T. Heinzl and A.I. Titov for inspiring questions and comments to be properly addressed in future studies.

\section*{References}
\bibliography{lphys16}
\end{document}